\documentclass{article}
\usepackage{opex3}
\begin{document}

%%%%%%%%%%%%%%%%%% title page information %%%%%%%%%%%%%%%%%%
\title{Atmosphere-like turbulence generation with surface-etched phase-screens}

\author{Stefan Hippler, Felix Hormuth, David J. Butler, Wolfgang Brandner, and Thomas Henning}

\address{Max-Planck-Institut f\"{u}r Astronomie \\ K\"{o}nigstuhl 17, D-69117 Heidelberg, Germany}

\email{hippler@mpia.de} %% email address is required

\homepage{http://www.mpia.de} %% author's URL, if desired


\begin{abstract}
We built and characterized an optical system that emulates the optical
characteristics of 
an 8m-class telescope like the Very Large Telescope. The system contains 
rotating glass phase-screens
to generate realistic atmosphere-like optical turbulence, as needed for testing multi-conjugate adaptive
optics systems.
In this paper we
present an investigation of the statistical properties of two phase-screens etched 
on glass-plate surfaces, obtained from Silios Technologies. 
Those etched screens are highly transmissive (above 85\%) from 0.45 to 2.5$\mu$m.  From direct imaging, their Fried parameter 
 r$_0$ values  (0.43$\pm$0.04\,mm and 0.81$\pm$0.03\,mm, respectively, at 0.633$\mu$m) agree
 with the expectation to within 10\%. 
 This is also confirmed by a comparison of measured and expected
 Zernike coefficient variances.
Overall, we find that those screens are quite reproducible, allowing
sub-millimetre r$_0$ values, which were difficult to achieve in the past.
We conclude that the telescope emulator and phase-screens form 
 a powerful atmospheric turbulence generator allowing 
 systematic testing of different kinds of AO instrumentation.
\end{abstract}

\ocis{(010.1080) Adaptive optics; (010.1330) Atmospheric turbulence} 
% REPLACE WITH CORRECT OCIS CODES FOR YOUR ARTICLE
%%% MAPS_OPEX_*.bbl is pasted below ...

%%%% MAPS_OPEX_*.bbl is pasted above...

%

\section{Introduction}
The optical image quality and in particular the angular resolution of ground
based telescopes is hampered by the refractive index variations of the Earth's atmosphere. These refractive index variations show a temporal and spatial behavior that under certain assumptions is well described by Kolmogorov turbulence \cite{Kolmogorov41, Tatarskii93, Conan95}.  

Adaptive optics (AO) instrumentation, available on almost all 8--10m-class telescopes, can compensate
for optical turbulence (seeing) with  certain restrictions. One of these restrictions is the isoplanatic angle, which is a result of using a single reference star for the AO and therefore compensating the integral effect of atmospheric turbulence in one direction only. Multi-conjugate AO systems that compensate the dominant turbulent layers instead, can increase the isoplanatic angle from a few tens of arcseconds \cite{Weiss02} to up to a few arcminutes at 2.2\,$\mu$m \cite{Berkefeld01}. 

With the advent of second generation AO systems like multi-conjugate
AO \cite{Marchetti03}, ground layer AO \cite{Tokovinin04},  extreme
AO \cite{Koehler04}, multiple field of view AO \cite{Ragazzoni02}, and multiple
object AO \cite{Neichel05}, it is  becoming more and more important to have a realistic 3D turbulence simulation tool available in the laboratory. Such a tool provides repeatable optical turbulence and therefore well defined atmospheric conditions. It can support the assembly, integration, and verification phase of novel AO instrumentation such that AO performance can be verified before the instrument is attached to the telescope.

In our previous papers \cite{Butler03,Butler04} we described the design,
manufacturing process, and laboratory characterization of ion-exchange
phase-screens. 
While such phase-screen technology allows repeatable 
optical turbulence generation over a wide spectral range,
0.5--2.5$\mu$m, it does not allow sub-millimetre r$_0$ values.
In this paper we focus on a new type of phase-screen,
based on transmissive glass substrates with aberrations imprinted on one surface. 
The motivation for investigating this new technique is this:
For laboratory characterization of astronomical adaptive optics systems
designed for D=8--40-m-class telescopes, D/r$_0$ values greater 30 are required.
As we can only use scaled-down telescope optics in a laboratory, with
typical values of D$_{lab}$=10\,mm, we need r$_0$ values smaller than
D$_{lab}$/30=1/3\,mm.
Glass phase screens with such small r$_0$ values were in the past difficult to
fabricate. 
We identified the surface etching technique as the only tool and
therefore best solution so far to this problem.

In the following, we firstly describe in
Sec.~\ref{maps} the design of MAPS, an optical set-up used to test the screens.
The design and manufacture of the screens is then outlined in
Sec.~\ref{fabrication}. Our analyses and results are presented
  in Sec.~\ref{charac}, and we briefly summarize our findings
 in Sec.~\ref{conclus}.

%present their design and
%measured statistical properties and compare the results with those obtained from ion-exchange phase-screens.

%

\section{MAPS, the multi atmospheric phase-screens and stars instrument for the visible and near-infrared}\label{maps}

MAPS is a laboratory tool that allows simulation of 3D atmospheric optical turbulence over a wide field of view up to 2 arcminutes. It consists of 3 main components. 
A plate for light sources, comprising 34 fiber connectors \cite{hippler06} to
 investigate wide field wavefront reconstruction for various artificial-star
  configurations. 
The second main part of the MAPS design \cite{Butler03} consists of two optical
tubes containing identical groups of lenses, 
fabricated by Janos Technology, Inc., Keene, NH, USA.
 The first tube collimates the point
sources of the reference plate and creates the telescope pupil. The
second tube refocuses the disturbed light beams into a 2
arcminute-wide focal plane with the optical characteristics of the
Very Large Telescope (VLT) f/15 Nasmyth focus.
The third main component of MAPS is a set of phase-screens. Up to 3
glass phase-screens can be mounted in between the two optical tubes. When
rotating, those screens emulate a turbulent atmosphere consisting of a ground or boundary layer, a
mid-altitude layer, and a high altitude layer. Each glass-plate can rotate
with an adjustable and reproducible speed to emulate different atmospheric wind speeds per
layer (see also \cite{Butler04}). 
Additionally, we can adjust the position of each glass-plate along the optical axis of the system to emulate atmospheric layers at different altitudes. Such a layered approach of atmospheric turbulence is supported by a number of experimental studies 
\cite{Klueckers98,Fuchs98,Weiss02,McKenna03,Avila06,Egner06b}.

The glass-plate phase-screen design was kindly provided by
ESO \cite{Kolb04}. The telescope pupil of MAPS is implemented
through a 13\,mm (equivalent to an 8-m telescope) pupil stop installed immediately behind the ground layer phase-screen when viewed from the direction of the collimating lens assembly.

% @@@@ Direct imaging tests at 632.8nm and 831.5nm show that  diffraction-limited
% @@@@ imaging is reached using this device over a @@@@@@ field-of-view. 
%

\section{Fabricating the surface etched phase-screens}\label{fabrication}
\subsection{Phase-screen phase maps}
\label{pixelsize}
\begin{figure}[t]
\begin{center}
\begin{tabular}{c}
\includegraphics[width=12cm]{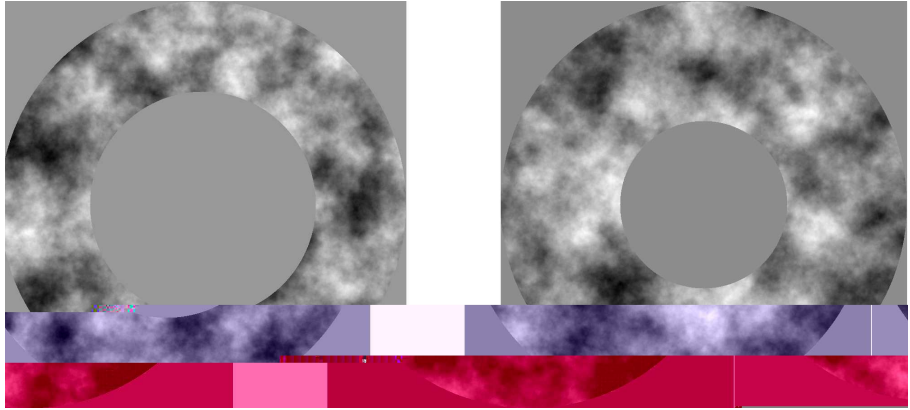}
\end{tabular}
\caption[phasemaps] 
%>>>> use \label inside caption to get Fig. number with \ref{}
{ \label{fig:pmaps} 
Phase maps of the ground layer phase-screen PS1 (left) and mid- to high-altitude
phase-screen PS2 (right).}
\end{center}
\end{figure}

The phase maps of the phase-screens manufactured for the MAPS assembly are
shown in Fig.~\ref{fig:pmaps}. In the following the screens will be denoted
as PS1 and PS2, with PS1 being the more turbulent screen. PS1 is used as
the ground layer screen in the final MAPS assembly, while PS2 will be used
as the mid- or high-layer screen. The peak-to-valley differences of
the phase maps are 7.9 and 5.5$\mu$m, respectively. One pixel of the
phase map corresponds to 0.1mm on the manufactured screen.

\subsection{Fused silica substrate}

\begin{figure}[ht]
\begin{center}
\begin{tabular}{c}
\includegraphics[width=12cm]{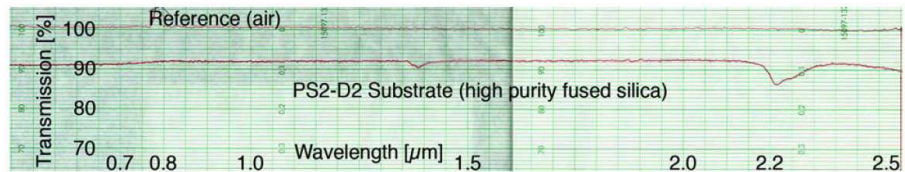}
\end{tabular}
\caption[glass_transmission] 
%>>>> use \label inside caption to get Fig. number with \ref{}
{ \label{fig:GLASS_TRANSMISSION} 
Transmission in \% as a function of wavelength, measured for phase-screen PS2.}
\end{center}
\end{figure}

%The phase profiles are etched between 32 to 64 levels. 
%The amplitude of the phase variations (peak to valley) reaches 8 microns. 
The glass-plate phase-screens from Silios Technologies, France, are manufactured with a wet etch process based on hydrochloric acid. The equipment is similar to standard semiconductor processing equipment used for 4-inch wafers. 
The phase pattern is etched into the 100\,mm diameter and 1.5\,mm thick glass
substrate consisting of fused silica (Corning code 7980) from Corning Inc.,
USA.  This high purity amorphous silicon dioxide has a very good transmission in the required spectral range from 0.5--2.5\,$\mu$m. As shown in Fig.~\ref{fig:GLASS_TRANSMISSION} the transmission is $>$90\% over the entire spectral range with only two small H$_2$O-absorption features at around 1.4 and 2.2\,$\mu$m. Visual 
inspection of the screens does not reveal any inhomogeneities or opaque
patches. The phase-screens are realized through a multilevel profile created
with either 5 (PS2) or 6 (PS1) masks, which eventually lead to 2$^5$=32 and 2$^6$=64 different levels of the phase map. 

\section{Characterization of glass-plate phase-screens from Silios
Technologies}\label{charac}

For the (statistical) characterization of the manufactured phase-screens we use
two different approaches: first, methods based on measurements
of the point spread function (PSF), and second, 
wavefront-based methods. The basic parameter we wanted
to measure and compare to the theo\-retical expectation, is the
well-known Fried parameter r$_0$. An equally important quantity
is the slope of the turbulence power spectrum or the slope
of the variance of the Zernike coefficients vs. mode number.
Although these quantities could be readily derived in the case
of the available computer-generated template wavefronts used for the
phase-screens, characterizing the power spectrum of the
manufactured screens requires an accurate measurement of the
actual distorted wavefront. 

\begin{figure}[t]
\begin{center}
        \includegraphics[width=12cm]{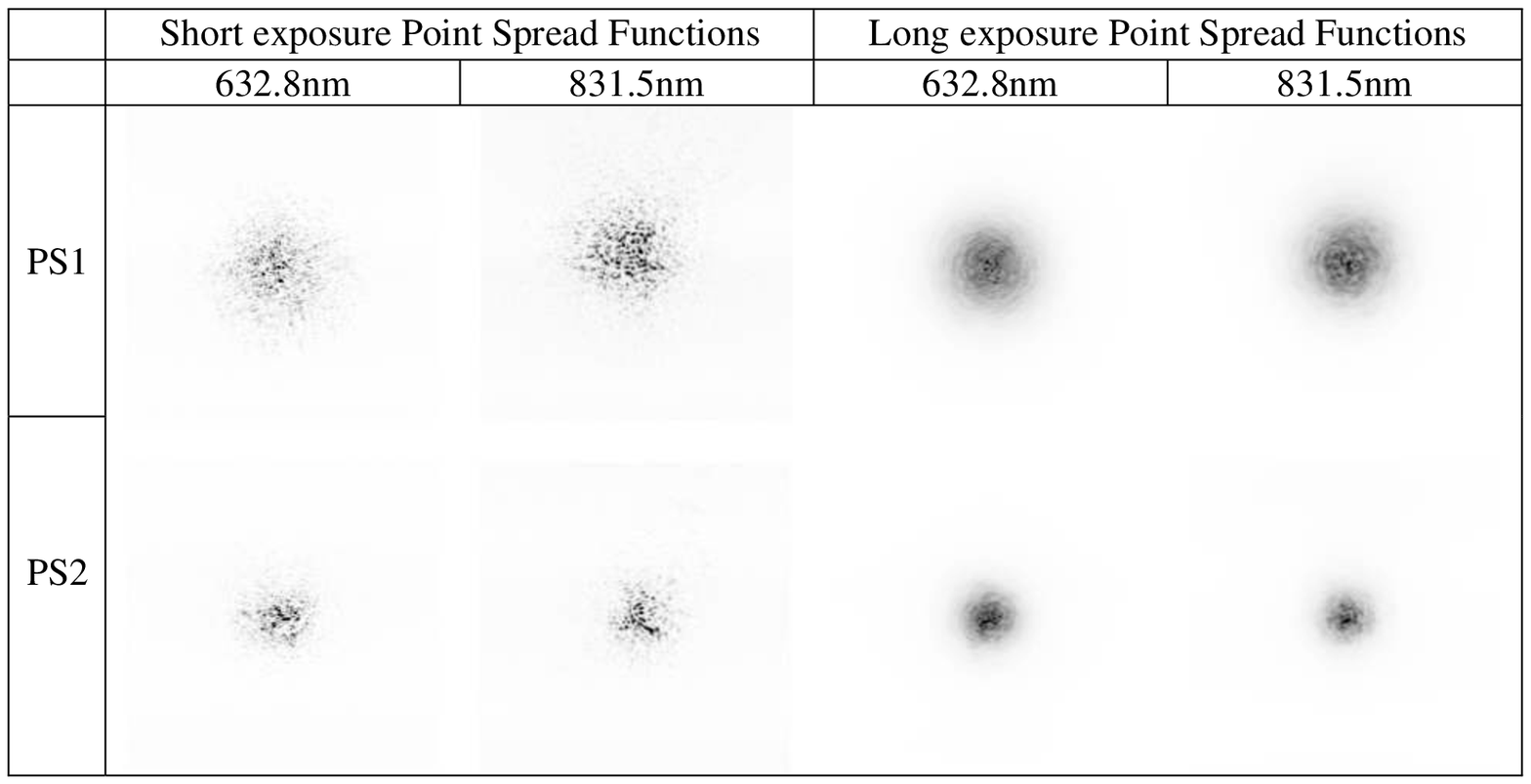}
        \caption{Four representative images of both short- and long-exposure point 
          spread functions  for PS1 and PS2, at 632.8nm
          and 831.5nm. The data was 
           taken using an aperture size of 13mm. 
          The size of each image  is 200x200 CCD pixel$^2$.
        \label{fig:psfexam}}
\end{center}
\end{figure}

\subsection{Point spread function (PSF) measurement}
To measure the PSF we put a single phase-screen at the ground
layer position in the MAPS assembly, using a 13mm pupil-stop as
the telescope aperture, which is equivalent to a real-world aperture of 8m. 
A single-mode fiber at the center of the
field of view on the input side of MAPS was used as light
source. Measurements were performed at two wavelengths, using
a HeNe laser with $\lambda$=632.8nm, and a diode laser with
$\lambda$=831.5nm.
A standard CCD with a pixel size of 6.9x6.9\,$\mu$m$^2$ was placed
in the focal plane of the imaging side to record PSF images.
%short exposure 
To capture short exposure PSFs, the phase-screen was rotated at a speed 
of typically 0.5 rpm, while the CCD acquired large sets of 
exposures (of the order of several thousand) with integration 
times of a few milliseconds. Afterwards, all short exposure images 
were stacked to generate the long exposure PSF image used for the
 coherence length (r$_0$) measurement.
In Fig. \ref{fig:psfexam} we show examples of short and long
exposure PSFs for both phase-screens at the two wavelengths.
In order to derive values for r$_0$ we additionally took 
images using pinholes of different
sizes as telescope aperture and recorded the resulting PSFs without
a phase-screen in the optical beam. 

The FWHM of the PSF obtained without
any turbulence -- the diffraction limited PSF -- is then given by:
\begin{equation}
  FWHM_{DL} = 1.02 \cdot \frac{\lambda}{D} \cdot S,
\end{equation}
where the subscript DL denotes the diffraction limited FWHM, and D is the
diameter of the pupil-stop defining the telescope aperture.
The factor S is the image scaling factor of our system, containing both focal 
length and pixel size and determined with the calibration images. 
Our calibration measurements are in agreement with the
predicted $\lambda/D$ behavior within the measurement errors.
For the FWHM measured in radians and the 
telescope diameter given in meters, S would be simply unity.
Knowing the value for S we are able to compute the
coherence length r$_0$ from the measured FWHM of our seeing limited long 
exposure PSFs (FWHM$_{SL}$) using the equation:
\begin{equation}
  r_0 = 0.98 \cdot \frac{\lambda}{FWHM_{SL}} \cdot S
\end{equation}

\begin{figure}[t]
\begin{center}
        \includegraphics[width=12cm]{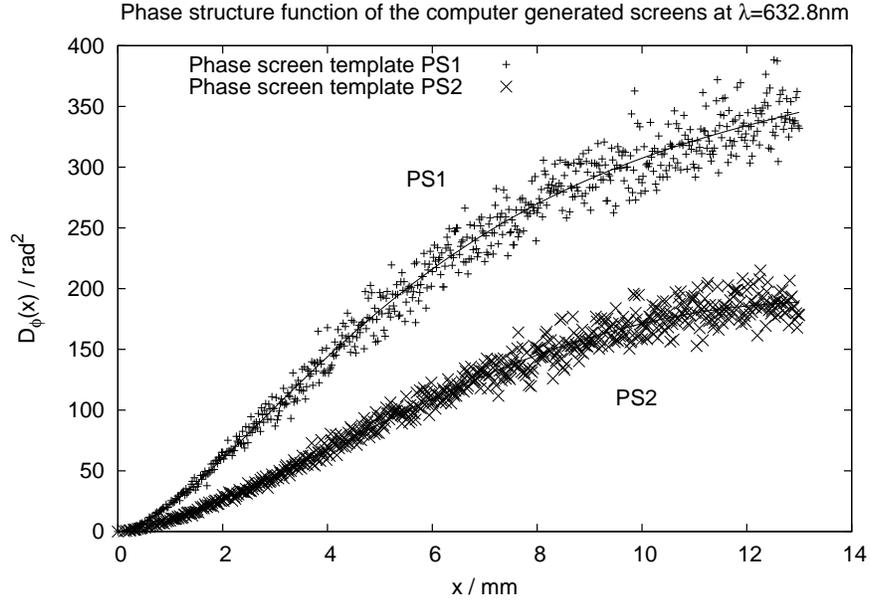}
        \caption{Phase structure function of the computer generated 
        screens for a wavelength of 632.8nm (optical phase differences scaled to this wavelength, 
        calculated on 360 randomly selected apertures of 13mm diameter with 
        10$^6$ point pairs in each). The solid lines represent a fit with L$_0$=22m and the values
        of r$_0$ as listed in Tbl.~\ref{tab:psfr0}
        \label{fig:phsf}}
\end{center}
\end{figure}

Before giving the actual results of these measurements, we first
describe how we derived  r$_0$  in the
case of the computer-generated
  template wavefronts. In a first step we reconstructed
the phase structure function described by
\begin{equation}
D_{\Phi}(x) = <[\Phi(x') - \Phi(x' + x)]^2>.
\end{equation}
The resulting structure functions of both screens are shown in Fig.~\ref{fig:phsf}. It is well visible that the structure function indeed flattens
out at larger values of x, clearly indicating a finite outer scale L$_0$. An outer scale of 
L$_0$=22m -- typical for the Paranal observatory -- was assumed for the design of the phase maps, and the measured flattening of the phase structure function is compatible with this value.

\begin{figure}[t]
\begin{center}
    \includegraphics[width=6.0cm]{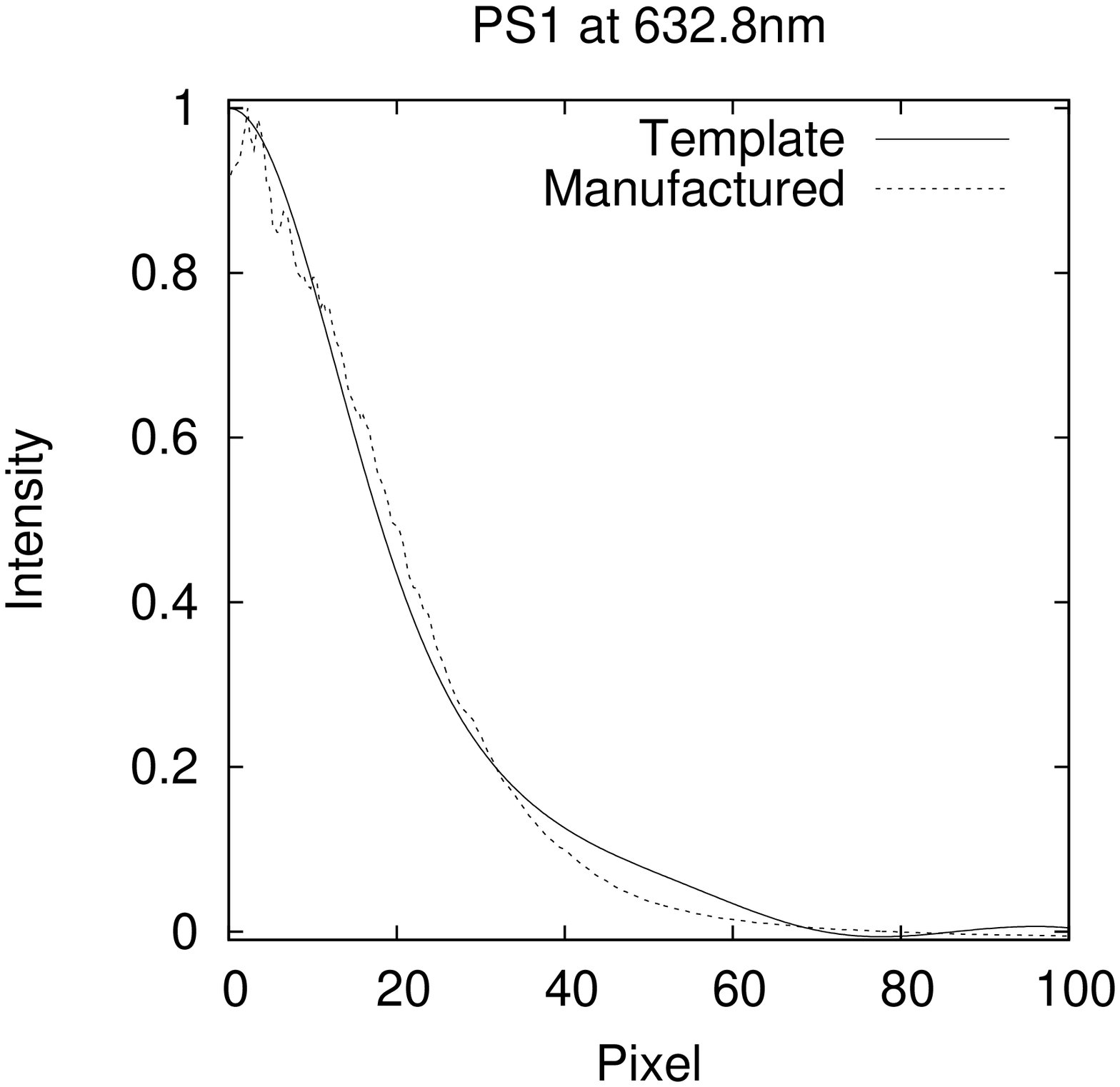}
    \includegraphics[width=6.0cm]{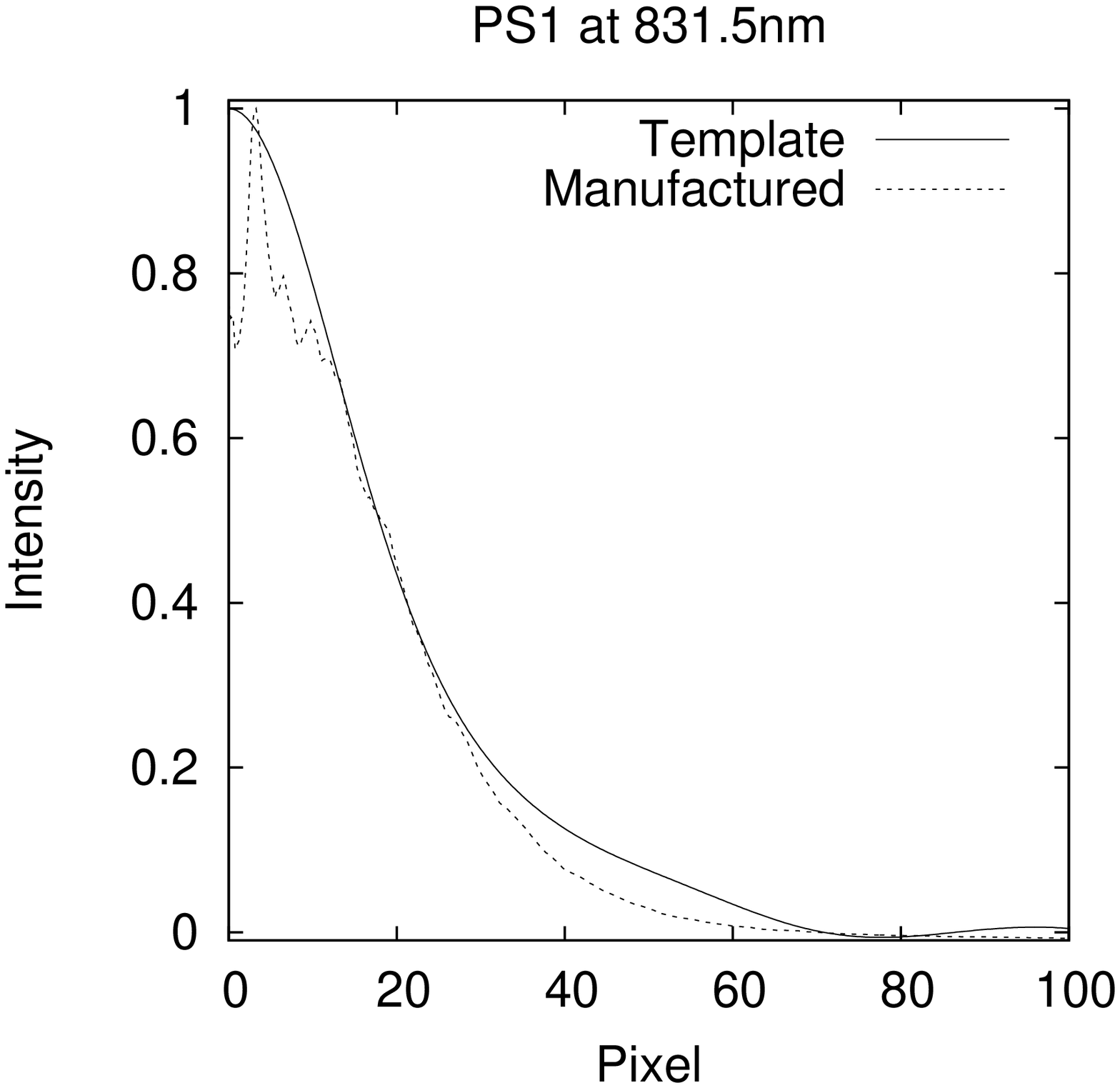}
    \includegraphics[width=6.0cm]{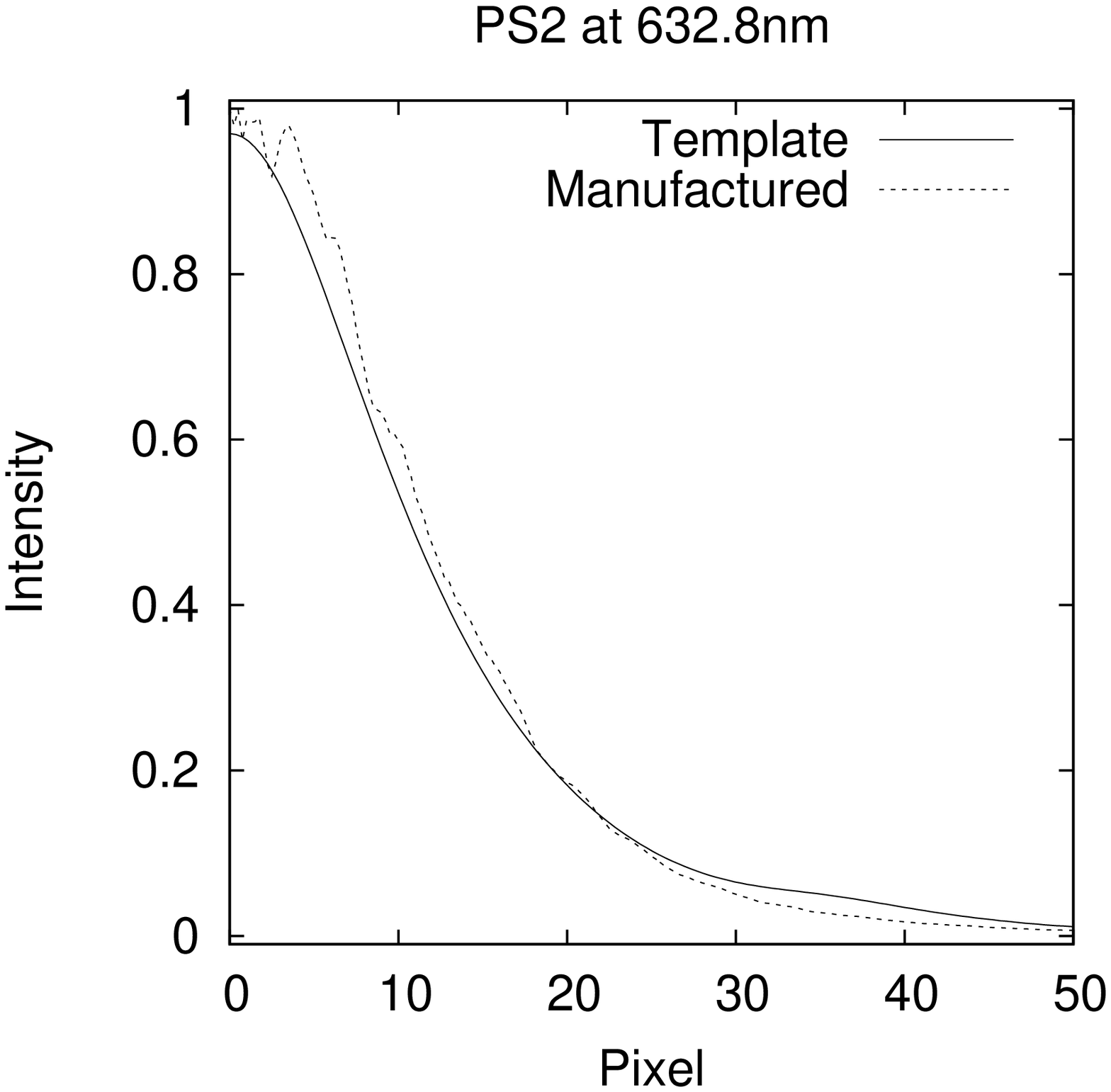}
    \includegraphics[width=6.0cm]{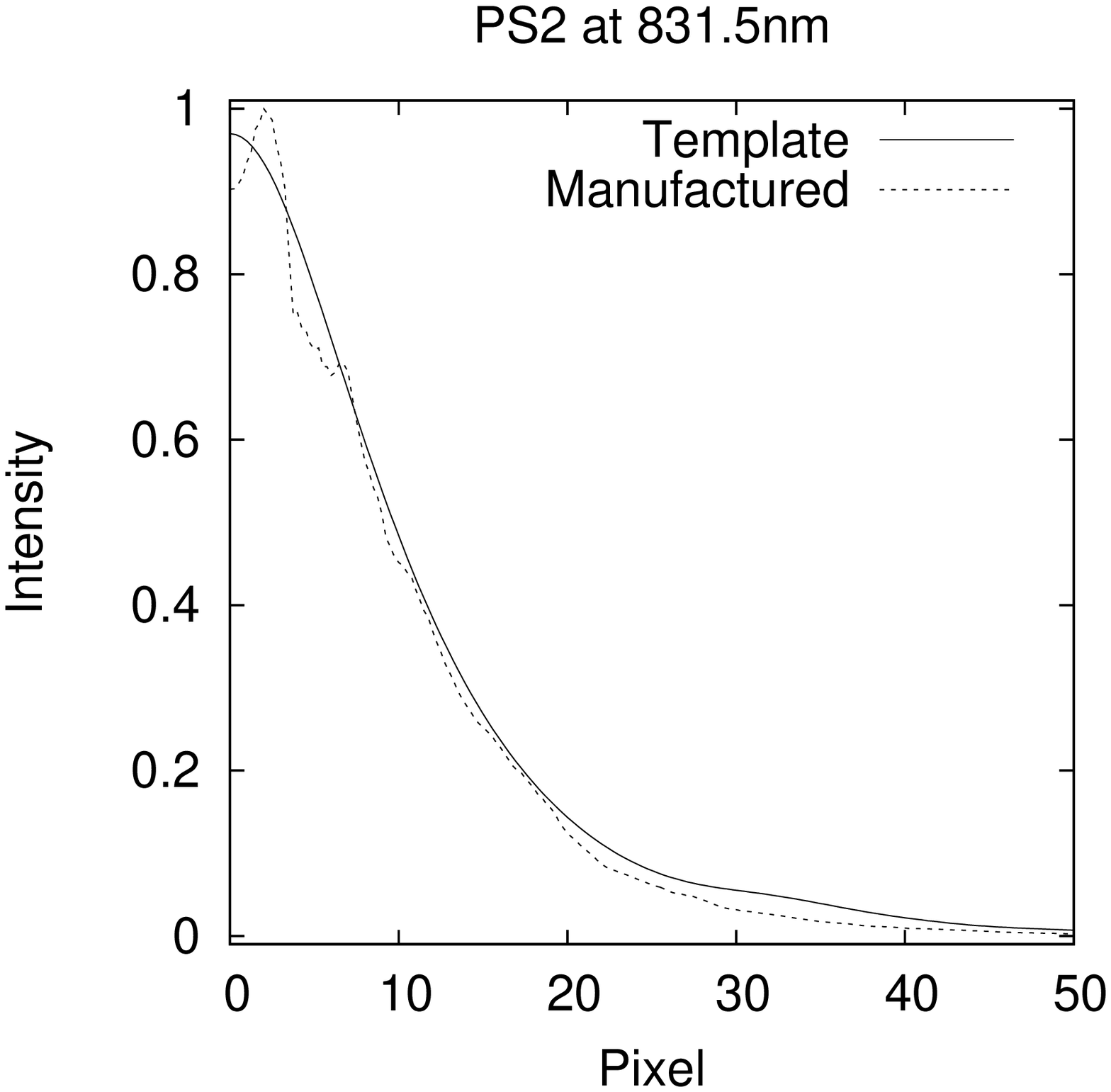}
  \caption{\label{fig:1dpsf} Demonstrating the close agreement between
 the expected (Template) and measured (Manufactured) long exposure PSFs
  for phase-screen PS1 (upper row) and PS2 (lower row). 
Shown are the normalized radial intensity profiles at   $\lambda$=632.8nm (left column) 
and $\lambda$=831.5nm (right column).
  Please note the different ranges of the  abscissa.}
\end{center}
\end{figure}

In a second step we used the phase structure function to compute
the optical transfer function (OTF) of the phase-screens using the equation
\begin{equation}
  OTF(x) = \exp[-0.5D_{\Phi}(x)].
\end{equation}
Since the PSF is the inverse Fourier transform of the OTF we are 
able to directly compute the  long exposure PSFs of
our phase-screen computer templates, and to compare them with our
experimental results for the manufactured screens. In Fig.~\ref{fig:1dpsf}
we show the theoretically expected PSFs, scaled to the measurement
wavelengths and the image scale of our set-up, together with the
radially averaged and normalized profiles of the PSFs measured 
from direct imaging.
From the FWHM of these profiles we derived the values given in 
Tbl.~\ref{tab:psfr0} for the coherence length r$_0$.

\begin{table}[ht]
\begin{center}
  \caption{Results of the r$_0$ measurement from direct imaging of the PSF.}
  \label{tab:psfr0}

  \begin{tabular}{lccc}
  \hline
  phase-screen             & r$_0(\lambda$=632.8nm)& r$_0(\lambda$=831.5nm)  & r$_0$@632.8nm /  r$_0$@831.5nm \\
  \hline
  PS1 computer template   &  0.44 $\pm$ 0.02mm  &     0.61 $\pm$ 0.02mm     &   0.72 $\pm$ 0.06 \\
  PS1 manufactured screen &  0.43 $\pm$ 0.04mm  &     0.63 $\pm$ 0.03mm     &   0.68 $\pm$ 0.10 \\
  \noalign{\smallskip}
  PS2 computer template   &  0.81 $\pm$ 0.03mm  &     1.12 $\pm$ 0.03mm     &   0.72 $\pm$ 0.05 \\
  PS2 manufactured screen &  0.73 $\pm$ 0.06mm  &     1.11 $\pm$ 0.04mm     &   0.66 $\pm$ 0.08 \\
  \hline
  \end{tabular}  
\end{center}
\end{table}

With the exception of the infrared measurement for phase-screen 2,
all measured values are in very good agreement with the theoretically expected
values.
The use of two different wavelengths also allows us to determine whether 
the wavelength dependence of r$_0$ on the phase-screen substrate differs
substantially from that expected in the Earth's atmosphere.
 Since the coherence length r$_0$ in the atmosphere scales
as $\lambda^{1.2}$, we can compare with the ratio of r$_0$ determined at 
  632.8 and 831.5nm with that for the atmosphere (i.e. 0.72).
All measurements are in agreement with the
theoretically expected value within one standard deviation.

\subsection{Point-spread functions demonstration videos obtained with MAPS}
%
% MOVIES
%
This section provides hyperlinks to three multimedia movies obtained with MAPS. The corresponding set-up parameters are listed in Tbl.~\ref{tab:maps_setup}. All videos were taken with a CCD camera positioned in the focal plane of MAPS. The videos nicely show the evolution of the atmosphere-like speckle pattern with time for various wind speeds and D/r$_0$ ratios. 
\begin{table}[t]
\begin{center}
  \caption{MAPS set-up parameters and MPEG demo movies.}
  \label{tab:maps_setup}
  \begin{tabular}{llll}
  \hline
  Set-up parameter & demo \#1 & demo \#2 & demo \#3 \\
  \hline
  Pupil diameter D & 13\,mm & 6.5\,mm & 6.5\,mm \\
  Wavelength & 633\,nm & 633\,nm & 633\,nm \\
  PS1 position & ground-layer & ground-layer & ground-layer \\
  PS1 wind speed & 1.5\,m/s  & 0.5\,m/s  & 1.5\,m/s \\
  PS2 position & high-layer &  high-layer  & high-layer \\
  PS2 wind speed & 3.0\,m/s & 1.0\,m/s & 3.0\,m/s \\
  Total D/r$_0$ & $\approx$40 & $\approx$20 & $\approx$20 \\
  Long exposure seeing & 0.6\,'' & 0.6\,'' & 0.6\,'' \\
  Speckle video frame rate & 4\, Hz & 4\,Hz & 4\,Hz \\
  MPEG video size & 2.3\,MB & 2.3\,MB & 2.3\,MB \\
%  Video snapshot & \includegraphics[width=2cm]{SpeckleMovie1.eps}
 % & \includegraphics[width=2cm]{SpeckleMovie2.eps} 
 % & \includegraphics[width=2cm]{SpeckleMovie3.eps} \\
%  Link to movie & \href{file:///Users/hippler/Desktop/Orlando2006/OPEX_PaperJuly2006/SpeckleMovie1.mpg}{Movie \#1}
%  & \href{file:///Users/hippler/Desktop/Orlando2006/OPEX_PaperJuly2006/SpeckleMovie2.mpg}{Movie \#2}
%  & \href{file:///Users/hippler/Desktop/Orlando2006/OPEX_PaperJuly2006/SpeckleMovie3.mpg}{Movie \#3} \\ 
  Link to movies & www.mpia.de & /homes/hippler/ & SpeckleMovieX.mpg X=1,2,3 \\

   \hline
  \end{tabular}  
\end{center}
\end{table}

\begin{figure}[ht]
\begin{center}
        \includegraphics[width=12cm]{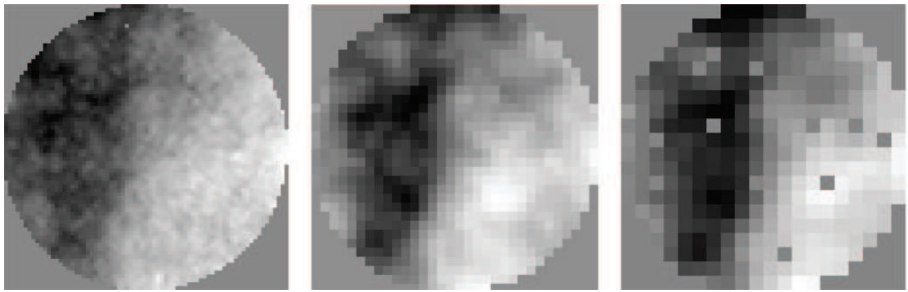}
        \caption{Examples of reconstructed phase maps of wavefronts 
              using the AOA Shack-Hartmann wavefront
          sensor for three lenslet sizes, namely 133$\mu$m, 
                 300$\mu$m, and 480$\mu$m (left to right).
The isolated bad pixels visible in the right panel result from dirty lenslets.
 %% The three images show
%                 the same location on the manufactured phase-screen PS1,
%                 using three different microlens arrays for the 
%                 measurement. Sub-aperture sizes were 133$\mu$m, 
%                 300$\mu$m, and 480$\mu$m, respectively (from left
%                 to right).
        \label{fig:wfmeas}}
\end{center}
\end{figure}

\subsection{Wavefront measurements}

The wavefronts produced by the manufactured phase-screens were measured\footnote{We also attempted to make similar wavefront measurements using a phase-shifting Twyman-Green interferometer from FISBA, Switzerland. This was unsuccessful as the wavefront of the 
phase-screens could not be reconstructed by the interferometer.}
with a commercial AOA Wavescope Shack-Hartmann sensor, using aperture
sizes of typically $\sim$10mm and microlens arrays with 133,
300, and 480$\mu$m microlens diameter (pitch size). The phase-screens were illuminated with
collimated light, using the HeNe laser and the infrared laser diode.
In Fig.~\ref{fig:wfmeas} we show the measured wavefront of the same location
on phase-screen PS1, measured with the three available microlens arrays.
\begin{figure}[h]
\begin{center}
        \includegraphics[width=12cm]{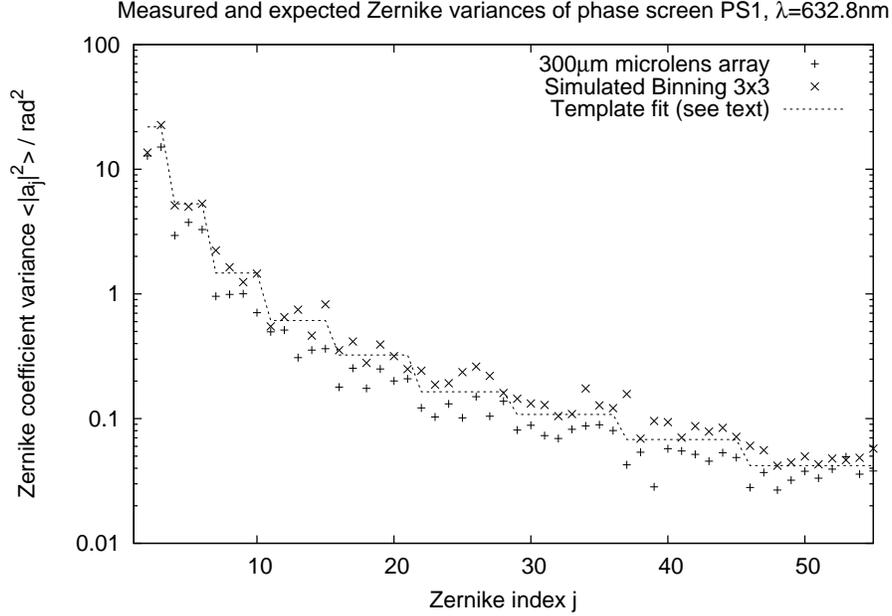}
        \caption{Zernike coefficient variances of phase-screen PS1 at
                 a wavelength of $\lambda=632.8\mu$m. 
                 Measurements are based on Shack-Hartmann wavefront gradients obtained with the 300$\mu$m-lenslet array.
                 The wavefronts of 100 randomly distributed test apertures with a
                 diameter of 10mm were reconstructed with a least-squares
                 Zernike decomposition. The plotted line (dashed) results from
                 a fit to the Zernike variances calculated for the
                 computer generated phase map under the assumption of
                 a finite outer scale of L$_0$=22m. Zernike mode
                 number 1 corresponds to the piston term and is
                 omitted in the plot. A Zernike decomposition of the computer generated phase-screen with a 3 by 3 pixel binning corresponding to a cell size of 300$\mu$m  is shown for comparison.
        \label{fig:zernike300}}
\end{center}
\end{figure} 
They show a broadly similar low spatial 
frequency structure, except for the left panel (i.e. using the 
133$\mu$m-array), which appears to differ markedly. While it 
shows considerable fine structure, it also appears to be somewhat flatter 
than the other pair of phase maps. This is not unexpected if we 
recall that on one hand the Silios phase-screens have a square
pixel-like structure, with each pixel, or cell, being ($s_{cell}$=) 
100$\mu$m-wide.  Yet, on the other hand, the Shack-Hartmann sensor only 
measures wavefront gradients, and that the minimum size of  each lenslet 
used needs to be $> 2 \cdot s_{cell}$. It is therefore 
not surprising that the 133$\mu$m-array appears to be relatively 
insensitive to the wavefront shape. 
% The 133$\mu$m-lenslet array is the
%ideal choice in terms of proper spatial sampling with respect to the finite pixel size of 100\,$\mu$m and
%the measured values of r$_0$.

\begin{figure}[t]
\begin{center}
        \includegraphics[width=6.0cm]{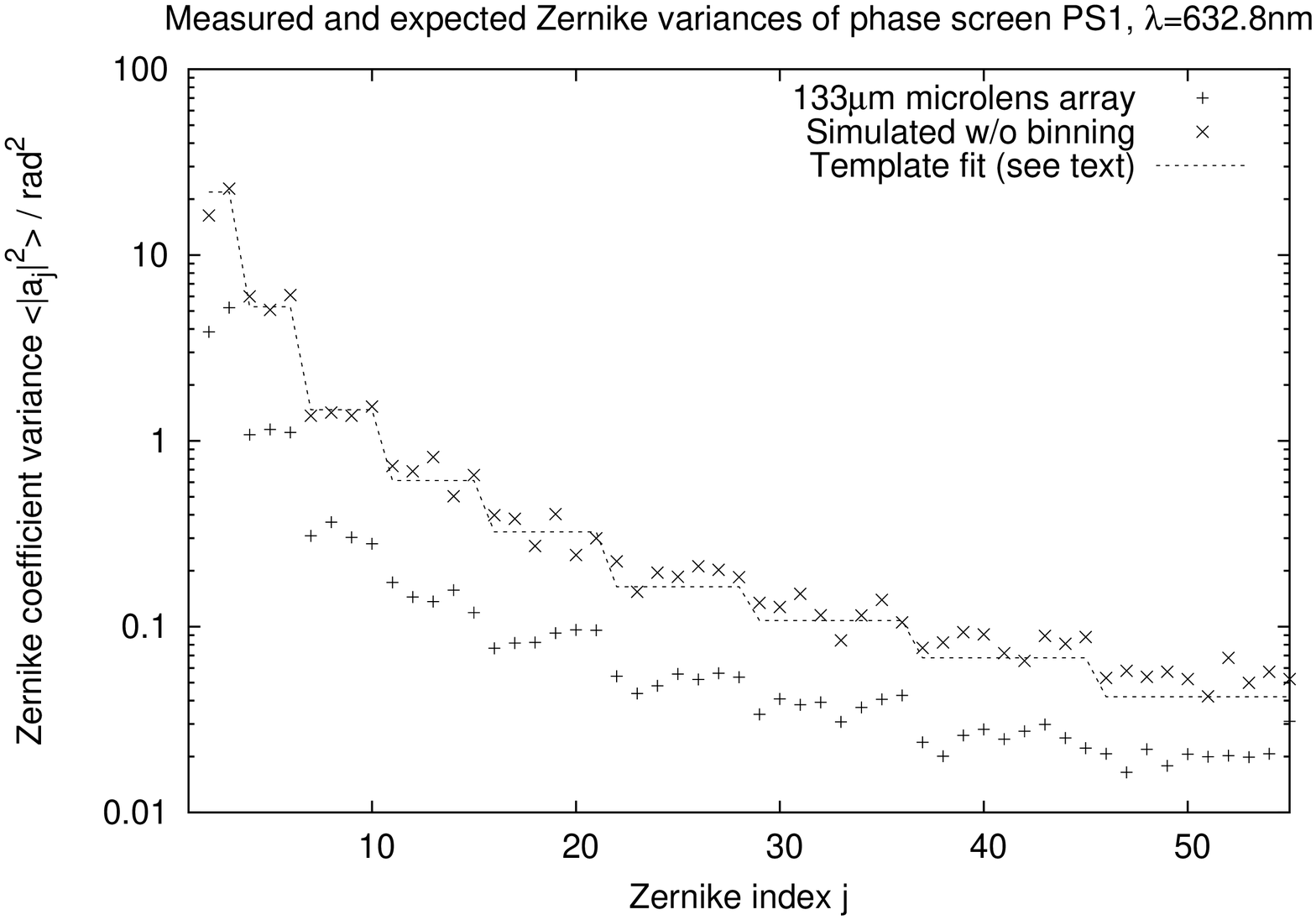}
         \includegraphics[width=6.0cm]{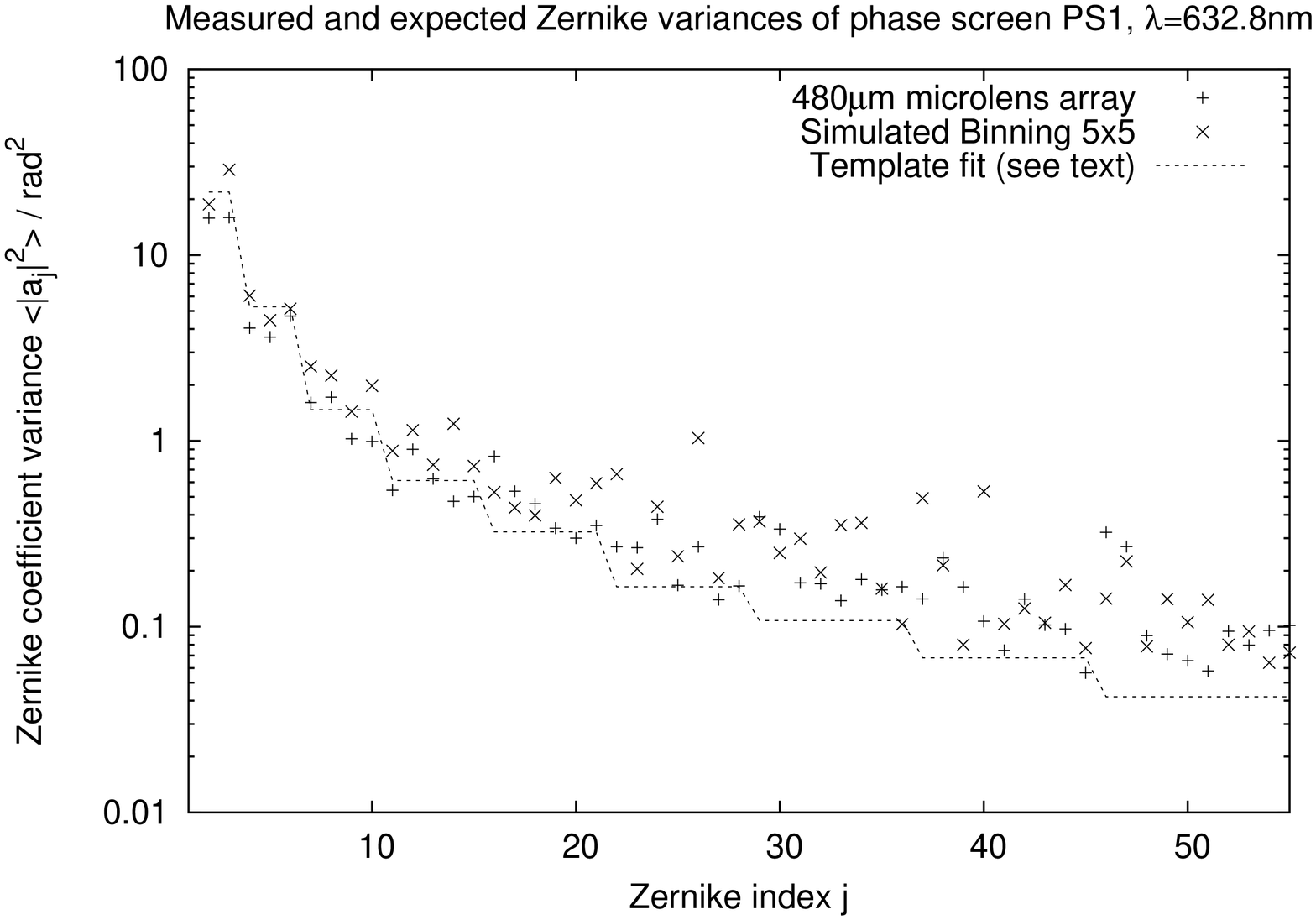}
        \caption{Left: same as Fig.~\ref{fig:zernike300}, but for the
         133$\mu$m-lenslet array. A Zernike decomposition of the original
         computer-generated phase-screen (i.e. without additional pixel
         binning),  corresponding to a cell
         size of 100$\mu$m, is shown for comparison. Right:  Also the same as
         Fig.~\ref{fig:zernike300}, but for the 480$\mu$m-lenslet array. A Zernike decomposition of the computer-generated phase-screen with a 5 by 5 pixel binning, corresponding to a cell size of 500$\mu$m, is shown for comparison.
         \label{fig:zernike100500}}
\end{center}
\end{figure}

In a quantitative approach, we decomposed both the measured
wavefronts and the computer-generated phase maps into their Zernike
representations and calculated the Zernike coefficient variances 
$<|a_j|^2>$. In the case of the computer generated phase maps,
the values were fitted accounting for a finite outer scale of L$_0$=22m.
The expected and measured Zernike variances are shown in Fig.\ref{fig:zernike300} and Fig.\ref{fig:zernike100500},
considering all three different microlens arrays. 
Indeed, the Zernike coefficients variance measurements confirm that the
133$\mu$m-lenslet array (lenslet size $< 2 \cdot s_{cell}$)  appears to be
  unsuitable for a statistical analysis of our phase-screens, as the
 variance estimated using the 133$\mu$m-array is considerably less 
than the expectation. As expected however the 
  300$\mu$m-array allows a better match to the
numerical simulation. The variance data for the
  480$\mu$m-array also provides a good match,
  except at Zernike indices above $\sim$20, where there is
increased scatter in $<|a_j|^2>$,  caused (at least in part) 
 by the (relatively)  small number of sampling elements
(lenslets) available for constraining the high spatial frequencies on the
phase-screen.

%
%\clearpage
\section{Conclusions}\label{conclus}

We have procured and characterized two phase-screens. We find that, within uncertainties, the statistical properties of both screens match those of the computer-generated templates. This suggests that the surface-etching technique allows reproducible phase-screens. In comparison to ion-exchange phase-screens  \cite{Butler03} we conclude that phase-screens with small r$_0$ values (strongest optical turbulence) are more reliably produced with the etching technology. We want to note that Shack-Hartmann wavefront measurements of phase-screens with a discrete pixel-like fine structure require a minimum
 Shack-Hartmann lenslet diameter; this diameter has to be at least twice as large as the
pixel size on the phase-screen.

\section*{Acknowledgments}
We thank Sebastian Egner for obtaining and providing the Speckle movies. It is a pleasure
to thank Eckhard Pitz for measuring the transmission of the phase-screens. Johann Kolb is 
thanked for providing all computer generated phase-screen data.
We thank the European Southern Observatory (ESO) 
for leading the procurement of Silios phase-screens.

%\bibliography{MAPS_REFS}   %>>>> bibliography data in MAPS_REFS.bib

\begin{thebibliography}{10}

\newcommand{\enquote}[1]{``#1''}
%\expandafter\ifx\csname url\endcsname\relax
%  \def\url#1{\texttt{#1}}\fi
%\expandafter\ifx\csname urlprefix\endcsname\relax\def\urlprefix{URL }\fi
%\providecommand{\eprint}[2][]{\url{#2}}

\bibitem{Kolmogorov41}
A.~Kolmogorov, \enquote{The local structure of turbulence in incompressible
  viscous fluids at very large Reynolds numbers,} Dokl. Akad. Nauk. SSSR
  \textbf{30}, 301--305 (1941). Reprinted in: S.K. Friedlander and L. Topper
  (editors). 1961. Turbulence: Classic Papers on Statistical Theory,
  Interscience Publications, New York, pp. 151-155.

\bibitem{Tatarskii93}
V.~I. {Tatarskii} and V.~U. {Zavorotny}, \enquote{{Atmospheric turbulence and
  the resolution limits of large ground-based telescopes: comment},} J. Opt.
  Soc. Am. A \textbf{10}, 2410--2414 (1993).

\bibitem{Conan95}
J.-M. {Conan}, G.~{Rousset}, and P.-Y. {Madec}, \enquote{{Wave-front temporal
  spectra in high-resolution imaging through turbulence},} J. Opt. Soc. Am. A
  \textbf{12}, 1559--1570 (1995).

\bibitem{Weiss02}
A.~{Wei{\ss}}, S.~{Hippler}, M.~{Kasper}, N.~{Wooder}, and J.~{Quartel},
  \enquote{{Simultaneous measurements of the Fried parameter $r_0$ and the
  isoplanatic angle $\theta_0$ using SCIDAR and adaptive optics - First
  results},} in \emph{ASP Conf. Ser. 266: Astronomical Site Evaluation in the
  Visible and Radio Range}, J.~{Vernin}, Z.~{Benkhaldoun}, and
  C.~{Mu{\~n}oz-Tu{\~n}{\'o}n}, eds., pp. 86--+ (2002).

\bibitem{Berkefeld01}
T.~{Berkefeld}, A.~{Glindemann}, and S.~{Hippler}, \enquote{{Multi-Conjugate
  Adaptive Optics with Two Deformable Mirrors - Requirements and Performance},}
  Exp. Astron. \textbf{11}, 1--21 (2001).

\bibitem{Marchetti03}
E.~{Marchetti}, N.~N. {Hubin}, E.~{Fedrigo}, J.~{Brynnel}, B.~{Delabre},
  R.~{Donaldson}, F.~{Franza}, R.~{Conan}, M.~{Le Louarn}, C.~{Cavadore},
  A.~{Balestra}, D.~{Baade}, J.-L. {Lizon}, R.~{Gilmozzi}, G.~J. {Monnet},
  R.~{Ragazzoni}, C.~{Arcidiacono}, A.~{Baruffolo}, E.~{Diolaiti},
  J.~{Farinato}, E.~{Vernet-Viard}, D.~J. {Butler}, S.~{Hippler}, and
  A.~{Amorin}, \enquote{{MAD the ESO multi-conjugate adaptive optics
  demonstrator},} in \emph{Adaptive Optical System Technologies II. Edited by
  Wizinowich, Peter L.; Bonaccini, Domenico. Proceedings of the SPIE, Volume
  4839, pp. 317-328 (2003).}, P.~L. {Wizinowich} and D.~{Bonaccini}, eds., pp.
  317--328 (2003).

\bibitem{Tokovinin04}
A.~{Tokovinin}, \enquote{{Seeing Improvement with Ground-Layer Adaptive
  Optics},} PASP \textbf{116}, 941--951 (2004).

\bibitem{Koehler04}
R.~{Koehler}, S.~{Hippler}, M.~{Feldt}, R.~{Gratton}, D.~{Gisler}, R.~{Stuik},
  and J.~{Lima}, \enquote{{Optimizing wavefront sensing for extreme AO},} in
  \emph{Advancements in Adaptive Optics. Edited by Domenico B. Calia, Brent L.
  Ellerbroek, and Roberto Ragazzoni. Proceedings of the SPIE, Volume 5490, pp.
  586-592 (2004).}, D.~{Bonaccini Calia}, B.~L. {Ellerbroek}, and
  R.~{Ragazzoni}, eds., pp. 586--592 (2004).

\bibitem{Ragazzoni02}
R.~{Ragazzoni}, E.~{Diolaiti}, J.~{Farinato}, E.~{Fedrigo}, E.~{Marchetti},
  M.~{Tordi}, and D.~{Kirkman}, \enquote{{Multiple field of view layer-oriented
  adaptive optics. Nearly whole sky coverage on 8 m class telescopes and
  beyond},} A\&A \textbf{396}, 731--744 (2002).

\bibitem{Neichel05}
B.~Neichel, T.~Fusco, M.~Puech, J.-M. Conan, M.~Lelouarn, E.~Gendron,
  F.~Hammer, G.~Rousset, P.~Jagourel, and P.~Bouchet, \enquote{{Adaptive Optics
  Concept For Multi-Objects 3D Spectroscopy on ELTs},} astro-ph
  \textbf{0512525} (2005).

\bibitem{Butler03}
D.~J. {Butler}, E.~{Marchetti}, J.~{B{ae}hr}, W.~{Xu}, S.~{Hippler}, M.~E.
  {Kasper}, and R.~{Conan}, \enquote{{Phase screens for astronomical
  multi-conjugate adaptive optics: application to MAPS},} in \emph{Adaptive
  Optical System Technologies II. Edited by Wizinowich, Peter L.; Bonaccini,
  Domenico. Proceedings of the SPIE, Volume 4839, pp. 623-634 (2003).}, P.~L.
  {Wizinowich} and D.~{Bonaccini}, eds., pp. 623--634 (2003).

\bibitem{Butler04}
D.~Butler, S.~Hippler, S.~Egner, W.~Xu, and J.~Baehr, \enquote{Broadband,
  static wave-front generation: Na-AG ion-exchange phase screens and telescope
  simulation,} Appl. Opt. \textbf{43}, 2813--2823 (2004).

\bibitem{hippler06}
S.~{Hippler}, F.~{Hormuth}, W.~{Brandner}, D.~{Butler}, T.~{Henning}, and
  S.~{Egner}, \enquote{{The MPIA multipurpose laboratory atmospheric turbulence
  simulator MAPS},} in \emph{Advances in Adaptive Optics II. Proceedings of the
  SPIE, Volume 6272} (2006).

\bibitem{Klueckers98}
V.~A. {Klueckers}, N.~J. {Wooder}, T.~W. {Nicholls}, M.~J. {Adcock},
  I.~{Munro}, and J.~C. {Dainty}, \enquote{{Profiling of atmospheric turbulence
  strength and velocity using a generalised SCIDAR technique},} A\&A Supplement
  \textbf{130}, 141--155 (1998).

\bibitem{Fuchs98}
A.~{Fuchs}, M.~{Tallon}, and J.~{Vernin}, \enquote{{Focusing on a Turbulent
  Layer: Principle of the ``Generalized SCIDAR''},} PASP \textbf{110}, 86--91
  (1998).

\bibitem{McKenna03}
D.~L. {McKenna}, R.~{Avila}, J.~M. {Hill}, S.~{Hippler}, P.~{Salinari}, P.~C.
  {Stanton}, and R.~{Weiss}, \enquote{{LBT facility SCIDAR: recent results},}
  in \emph{Adaptive Optical System Technologies II. Edited by Wizinowich, Peter
  L.; Bonaccini, Domenico. Proceedings of the SPIE, Volume 4839, pp. 825-836
  (2003).}, P.~L. {Wizinowich} and D.~{Bonaccini}, eds., pp. 825--836 (2003).

\bibitem{Avila06}
R.~{Avila}, E.~{Carrasco}, F.~{Iba{\~n}ez}, J.~{Vernin}, J.-L. {Prieur}, and
  D.~X. {Cruz}, \enquote{{Generalized SCIDAR Measurements at San Pedro
  M{\'a}rtir. II. Wind Profile Statistics},} PASP \textbf{118}, 503--515
  (2006).

\bibitem{Egner06b}
S.~Egner, E.~Mascidari, D.~McKenna, T.~M. Herbst, and W.~Gaessler,
  \enquote{{G-SCIDAR measurements on Mt.~Graham: recent results},} in
  \emph{Advances in Adaptive Optics II. Proceedings of the SPIE, Volume 6272}
  (2006).

\bibitem{Kolb04}
J.~{Kolb}, E.~{Marchetti}, S.~{Tisserand}, F.~{Franza}, B.~{Delabre},
  F.~{Gonte}, R.~{Brast}, S.~{Jacob}, and F.~{Reversat}, \enquote{{MAPS: a
  turbulence simulator for MCAO},} in \emph{Advancements in Adaptive Optics.
  Edited by Domenico B. Calia, Brent L. Ellerbroek, and Roberto Ragazzoni.
  Proceedings of the SPIE, Volume 5490, pp. 794-804 (2004).}, D.~{Bonaccini
  Calia}, B.~L. {Ellerbroek}, and R.~{Ragazzoni}, eds., pp. 794--804 (2004).

\end{thebibliography}
\bibliographystyle{osajnl}   %>>>> makes bibtex use osajnl.bst

\end{document}